**Thermodynamics of amide + amine mixtures. 2. Volumetric, speed of sound and refractive index data for *N,N*-dimethylacetamide + *N*-propylpropan-1-amine, + *N*-butylbutan-1-amine, + butan-1-amine, or + hexan-1-amine systems at several temperatures**

Fernando Hevia, Ana Cobos, Juan Antonio González*, Isaías García de la Fuente and Víctor Alonso

G.E.T.E.F., Departamento de Física Aplicada, Facultad de Ciencias, Universidad de Valladolid, Paseo de Belén, 7, 47011 Valladolid, Spain.

*e-mail: jagl@termo.uva.es; Fax: +34-983-423136; Tel: +34-983-423757

**Abstract**

Data on density, $\rho$, speed of sound, $c$, and refractive index, $n_D$, of binary systems containing *N,N*-dimethylacetamide (DMA) + *N*-propylpropan-1-amine (DPA) or + butan-1-amine (BA) at 293.15 K, 298.15 K and 303.15 K, and + *N*-butylbutan-1-amine (DBA) or + hexan-1-amine (HxA) at 298.15 K are reported. A densimeter and sound analyzer Anton Paar DSA 5000 has been used for the measurement of $\rho$ and $c$, whereas $n_D$ values have been obtained by means of a refractometer RFM970 from Bellingham+Stanley. Also, values of excess molar volumes, $V_m^E$, excess isentropic compressibilities, $\kappa_S^E$, excess speeds of sound, $c^E$, excess isobaric thermal expansion coefficients, $\alpha_p^E$, and of excess refractive indices, $n_D^E$, have been determined from these data. The investigated systems are characterized by amide-amine interactions and structural effects, as it is shown by their negative or low positive $V_m^E$ values and by the results from the application of the Prigogine-Flory-Patterson (PFP) model. The breaking of amine-amine interactions is more relevant in systems containing linear primary amines than in those with linear secondary amines, and the $V_m^E$ values are lower for the latter systems. Molar refraction has been used to evaluate the dispersive interactions in the mixtures under study, yielding the result that DPA and HxA systems present similar dispersive interactions and mainly differ in their dipolar character. Steric hindrance of the amide group in DMA leads to weaker amide-amine interactions than in the corresponding *N,N*-dimethylformamide (DMF) + amine systems.

# 1. Introduction

*N,N*-dimethylformamide (DMF) and *N,N*-dimethylacetamide (DMA) are very polar compounds (their dipole moment is 3.7 D [1, 2]) widely used in the industry, since they are aprotic protophilic substances with excellent donor-acceptor properties and solubility. In addition, they are employed for the separation of aromatic compounds and petroleum hydrocarbons. Amides are very common in nature and are found in proteins, RNA, DNA, amino acids, hormones and vitamins. The knowledge of liquid mixtures containing the amide functional group is necessary for a deeper understanding of more complex molecules, as those of biological interest [3]. Moreover, amides deserve to be investigated, as in pure state they show a significant local order [4]. In the case of *N,N*-dialkylamides, due to the absence of hydrogen bonds, this has been attributed to the existence of strong dipolar interactions [5].

Linear primary and secondary amines can form hydrogen bonds, appearing self-associated complexes and even heterocomplexes in mixtures with other associated compounds [6-8]. The amine group is also present in compounds of great biological significance. The proteins usually bound to DNA polymers contain various amine groups [9]. Histamine and dopamine are amines with the role of neurotransmitters [8, 10], and the breaking of amino acids releases amines. On the other hand, the ions of many ionic liquids used in technical applications are related to amines [11].

In earlier works, we have studied the thermodynamic properties of mixtures containing ketones and amines [12-19]. It is interesting to examine the effect of replacing a ketone, a moderately polar compound, by a more polar one, such as an amide. In our previous study [20], we have reported data on density, $\rho$, speed of sound, $c$, and refractive index, $n_\text{D}$ of the binary systems DMF + *N*-propylpropan-1-amine (DPA) or + butan-1-amine (BA) at (293.15-303.15) K, and + *N*-butylbutan-1-amine (DBA) or + hexan-1-amine (HxA) at 298.15 K. Now, we continue this series of works by replacing DMF by DMA, and treating these systems by means of the Prigogine-Flory-Patterson (PFP) model [21]. A survey of literature data shows that there are no experimental data on the considered mixtures. Nevertheless, DMF, or DMA + aniline or pyridine mixtures have been investigated rather extensively, reporting calorimetric, volumetric, vapor-liquid equilibria, $c$, or $n_\text{D}$ data [22-27]. Interestingly, at equimolar composition and 298.15 K, the excess molar enthalpies ($H_\text{m}^\text{E}$) of the DMF or DMA + aniline systems are, respectively, –2946 J·mol$^{-1}$ [25] and –352 J·mol$^{-1}$ [27], which underlines the importance of interactions between unlike molecules in such systems.

## 2. Experimental

*Materials*

Table 1 contains information about the source and the purity of the compounds, which have been used without further purification. Table 2 lists experimental values of $\rho$, $c$, $n_\mathrm{D}$, thermal expansion coefficient, $\alpha_p$, isentropic compressibility, $\kappa_S$, and isothermal compressibility, $\kappa_T$, for the pure compounds. Our values are in good agreement with the literature data.

*Apparatus and procedure*

Binary mixtures have been prepared by mass in small vessels of about 10 cm$^3$, using an analytical balance HR-202 (weighing accuracy 0.01 mg), with all weighings corrected for buoyancy effects. The standard uncertainty in the final mole fraction is estimated to be 0.0001. Molar quantities were calculated using the relative atomic mass Table of 2015 issued by the Commission on Isotopic Abundances and Atomic Weights (IUPAC) [28].

Temperatures were measured using Pt-100 resistances, calibrated according to the ITS-90 scale of temperature, against the triple point of water and the melting point of Ga. The standard uncertainty of the equilibrium temperature measurements is 0.01 K and 0.02 K for $\rho$ and $n_\mathrm{D}$ measurements, respectively.

Densities and speeds of sound have been measured using a vibrating-tube densimeter and sound analyzer DSA 5000 from Anton Paar, which is automatically thermostated within 0.01 K. The calibration of the device has been described in a previous work [14]. The repeatability of the $\rho$ measurements is 0.005 kg·m$^{-3}$, whereas their overall standard uncertainty is 1·10$^{-2}$ kg·m$^{-3}$. The determination of the speed of sound is based on the measurement of the time of propagation of short acoustic pulses, whose central frequency is 3 MHz [29], and which are transmitted repeatedly through the sample. The repeatability of these $c$ measurements is 0.1 m·s$^{-1}$ and their standard uncertainty is 0.2 m·s$^{-1}$. The excess volume, $V_\mathrm{m}^\mathrm{E}$, and the excess speed of sound, $c^\mathrm{E}$, of the system cyclohexane + benzene have been measured at (293.15-303.15) K to check the experimental technique. The experimental results and published values [30-32] are in good agreement. The standard uncertainty of $V_\mathrm{m}^\mathrm{E}$ is (0.010 $|V_\mathrm{m,max}^\mathrm{E}|$ + 0.005 cm$^3$·mol$^{-1}$), where $|V_\mathrm{m,max}^\mathrm{E}|$ stands for the maximum absolute experimental value of $V_\mathrm{m}^\mathrm{E}$ respect to the composition. The standard uncertainty of $c^\mathrm{E}$ is estimated to be 0.4 m·s$^{-1}$.

A refractometer RFM970 from Bellingham+Stanley has been used for the $n_\mathrm{D}$ measurements. The technique is based on the optical detection of the critical angle at the wavelength of the sodium D line (589.3 nm). The temperature is controlled by means of Peltier modules and its stability is 0.02 K. The refractometer has been calibrated using 2,2,4-trimethylpentane and toluene at the working temperatures (293.15-303.15) K, as recommended by Marsh [33]. The repeatability of the measurements is 0.00004, and the standard uncertainty is 0.00008.

## 3. Equations

The experimental values of $\rho$, molar volume, $V_m$, $\alpha_p$, and $\kappa_S$, can be obtained by means of a densimeter and sound analyzer rather directly. The values of $\alpha_p = -(1/\rho)(\partial \rho/\partial T)_p$ have been calculated under the assumption that $\rho$ depends linearly on $T$ in the range of temperatures considered. Moreover, as long as it is possible to neglect the dispersion and absorption of the acoustic wave, $\kappa_S$ can be determined using $\rho$ and $c$ values through the Newton-Laplace equation:

$$\kappa_S = \frac{1}{\rho c^2} \tag{1}$$

The values $F^{id}$ of a quantity, $F$, for an ideal mixture at the same temperature and pressure as the investigated solution are calculated from the relations:

$$F^{id} = x_1 F_1^* + x_2 F_2^* \qquad (F = V_m, C_{pm}) \tag{2}$$

$$F^{id} = \phi_1 F_1^* + \phi_2 F_2^* \qquad (F = \alpha_p, \kappa_T) \tag{3}$$

where $F_i^*$ denotes the property for the pure component $i$, $C_{pm}$ is the molar heat capacity at constant pressure, $\kappa_T$ is the isothermal compressibility and $\phi_i = x_i V_{mi}^* / V_m^{id}$ represents the ideal volume fraction. In the case of $\kappa_S$ and $c$, the following expressions are used:

$$\kappa_S^{id} = \kappa_T^{id} - \frac{T V_m^{id} (\alpha_p^{id})^2}{C_{pm}^{id}} \tag{4}$$

$$c^{id} = \left(\frac{1}{\rho^{id} \kappa_S^{id}}\right)^{1/2} \tag{5}$$

being $\rho^{id} = (x_1 M_1 + x_2 M_2)/V_m^{id}$ the ideal density, and $M_i$ the molar mass of the pure component $i$. For the refractive index, $n_D$, the ideal values are obtained from the equation [34]:

$$n_D^{id} = \left[\phi_1 (n_{D1}^*)^2 + \phi_2 (n_{D2}^*)^2\right]^{1/2} \tag{6}$$

The excess properties, $F^E$, are then obtained from the relation:

$$F^{\mathrm{E}} = F - F^{\mathrm{id}} \qquad \left(F = V_{\mathrm{m}}, \kappa_S, c, \alpha_p, n_{\mathrm{D}}\right) \tag{7}$$

## 4. Results

Values of $\rho$, $c$, and $V_{\mathrm{m}}^{\mathrm{E}}$ as functions of $x_1$, the mole fraction of DMA, and at the considered temperatures are included in Table 3. For DBA or HxA mixtures, the measurements were made at 298.15 K only, due to: (i) their low $|V_{\mathrm{m}}^{\mathrm{E}}|$ values; (ii) the weak temperature dependence of $V_{\mathrm{m}}^{\mathrm{E}}$ encountered for the systems with BA or DPA. The corresponding results of $\kappa_S^{\mathrm{E}}$, $c^{\mathrm{E}}$, and $\alpha_p^{\mathrm{E}}$ at 298.15 K are given in Table 4. The $n_{\mathrm{D}}$ values and their corresponding excess functions, $n_{\mathrm{D}}^{\mathrm{E}}$, are collected in Table 5. Our experimental method is not accurate enough to determine $n_{\mathrm{D}}^{\mathrm{E}}$ values for the systems containing DBA or HxA. Some of these results are represented in Figures 1-7. We have not found literature data for comparison.

The data have been fitted by an unweighted linear least-squares regression to a Redlich-Kister equation [35]:

$$F^{\mathrm{E}} = x_1 (1 - x_1) \sum_{i=0}^{k-1} A_i (2x_1 - 1)^i \qquad \left(F = V_{\mathrm{m}}, \kappa_S, c, \alpha_p, n_{\mathrm{D}}\right) \tag{8}$$

For each system and property, the number, $k$, of necessary coefficients for this regression has been determined by applying an F-test of additional term [36] at 99.5% confidence level. Table 6 includes the parameters $A_i$ obtained, and the standard deviations $\sigma(F^{\mathrm{E}})$, defined by:

$$\sigma(F^{\mathrm{E}}) = \left[\frac{1}{N-k} \sum_{j=1}^{N} \left(F_{\mathrm{cal},j}^{\mathrm{E}} - F_{\mathrm{exp},j}^{\mathrm{E}}\right)^2\right]^{1/2} \tag{9}$$

where the index $j$ takes one value for each of the $N$ experimental data $F_{\mathrm{exp},j}^{\mathrm{E}}$, and $F_{\mathrm{cal},j}^{\mathrm{E}}$ is the corresponding value of the excess property $F^{\mathrm{E}}$ calculated from equation (8).

## 5. Prigogine-Flory-Patterson model

In this version of the Flory theory, the excess volumes can be expressed as the sum of three terms [21]: an interactional contribution, proportional to $\chi_{12}$ (the interactional Flory parameter); a free volume contribution (the so-called curvature term), related to the difference in the degree of thermal expansion between the two components, and a $p^*$ contribution which arises from the differences in the internal pressures of the components. The mentioned terms are given, respectively, by:

$$\frac{V_{m,\text{interac}}^{E}}{x_1 V_{m,1}^* + x_2 V_{m,2}^*} = \frac{(\bar{V}^{1/3}-1)\bar{V}^{2/3}\Psi_1 \theta_2 (\chi_{12}/p_1^*)}{\frac{4}{3}\bar{V}^{-1/3}-1} \qquad (10)$$

$$\frac{V_{m,\text{curvature}}^{E}}{x_1 V_{m,1}^* + x_2 V_{m,2}^*} = -\frac{(\bar{V}_1-\bar{V}_2)^2 \left(\frac{14}{9}\bar{V}^{-1/3}-1\right)\Psi_1 \Psi_2}{\left(\frac{4}{3}\bar{V}^{-1/3}-1\right)\bar{V}} \qquad (11)$$

$$\frac{V_{m,p^*\text{effect}}^{E}}{x_1 V_{m,1}^* + x_2 V_{m,2}^*} = \frac{(\bar{V}_1-\bar{V}_2)(p_1^*-p_2^*)\Psi_1 \Psi_2}{p_2^* \Psi_1 + p_1^* \Psi_2} \qquad (12)$$

In these equations the contact energy fraction $\Psi_i$ is defined by:

$$\Psi_i = \frac{\varphi_i p_i^*}{\varphi_1 p_1^* + \varphi_2 p_2^*} \qquad (13)$$

The remaining symbols have their usual meaning [37-39]. $\bar{V} = V_m / V_m^*$ and $\bar{V}_i = V_{m,i}/V_{m,i}^*$ are the reduced volume of the mixture and of component $i$, respectively; $V_{m,i}^*$, $p_i^*$ and $T_i^*$ are the characteristic parameters (reduction parameters) of the pure liquids which are obtained from experimental data, such as $\alpha_{p,i}$ and $\kappa_{T,i}$. For mixtures, the corresponding parameters are calculated as follows [38, 39]:

$$V_m^* = x_1 V_{m,1}^* + x_2 V_{m,2}^* \qquad (14)$$

$$T^* = \frac{\varphi_1 p_1^* + \varphi_2 p_2^* - \varphi_1 \theta_2 \chi_{12}}{\frac{\varphi_1 p_1^*}{T_1^*} + \frac{\varphi_2 p_2^*}{T_2^*}} \qquad (15)$$

$$p^* = \varphi_1 p_1^* + \varphi_2 p_2^* - \varphi_1 \theta_2 \chi_{12} \qquad (16)$$

Finally, $\varphi_i = x_i V_{m,i}^* / \sum x_j V_{m,j}^*$ is the segment fraction and $\theta_2$ is site fraction ($=\varphi_2/(\varphi_2 + S_{12}\varphi_1)$). $S_{12}$ is the so-called geometrical parameter of the mixture, which, assuming that the molecules are spherical, is calculated as $S_{12} = \left(V_{m,1}^*/V_{m,2}^*\right)^{-1/3}$.

### 5.1. Theoretical results

Table 7 lists the values of $V_{m,i}^*$ and $p_i^*$ used in this work. $\chi_{12}$ values determined from $V_m^E$ at 298.15 K and equimolar composition are given in Table 8, which also contains the different contributions to $V_m^E$ calculated according to equations (10)-(12). A comparison between experimental and theoretical results is shown, for some selected mixtures, in Figures 8 and 9.

## 6. Discussion

In the present section, the values of the thermophysical properties and the excess functions are referred to $T = 298.15$ K and $x_1 = 0.5$.

DMA is a strongly polar compound (dipole moment $\mu / D = 3.7$ [1]). This is reflected in the fact that DMA + alkane mixtures present miscibility gaps up to quite high temperatures. For instance, the upper critical solution temperature of the heptane system is 309.8 K [40].

The amines considered in this work are linear, either primary or secondary. They are weakly self-associated and their dipole moments $\mu / D$ are low: 1.3 (BA) [41], 1.3 (HxA) [2], 1.0 (DPA) [41], and 1.1 (DBA) [41]. The values of the excess molar enthalpy, $H_m^E / \text{J·mol}^{-1}$, for the heptane mixtures are: 1192 (BA) [42], 962 (HxA) [42], 424 (DPA) [43], and 317 (DBA) [43]. These values can be explained in terms of the breaking of amine-amine interactions upon mixing. We note that $H_m^E$ values are lower for systems with secondary amines, as the amine group is more sterically hindered and self-association is lower in such amines. The corresponding values of $V_m^E(\text{heptane}) / \text{cm}^3 \cdot \text{mol}^{-1}$ are: 0.7171 (BA) [44], 0.3450 (HxA) [44], 0.2752 (DPA) [45], and 0.0675 (DBA) [45]. It is well stated that positive $V_m^E$ values arise from the disruption of interactions between like molecules, whereas negative ones appear when interactions between unlike molecules are created and/or when structural effects (differences in size and shape [46-48] or interstitial accommodation [49]) exist. The parallel change of $H_m^E$ and $V_m^E$ indicates that the disruption of amine-amine interactions upon mixing is the main contribution to $V_m^E$. Nevertheless, the low value of $V_m^E$ in the DBA + heptane system and the negative one of the DBA + hexane system, –0.1854 cm$^3$·mol$^{-1}$ [50], allow to state that structural effects are present, since this is suggested to be the most relevant contribution when a positive $H_m^E$ value is together with a negative $V_m^E$ value [48].

For the DMA + amine mixtures, we have obtained here either negative or small and positive $V_m^E / \text{cm}^3 \cdot \text{mol}^{-1}$ values (Figures 1, 2): –0.1940 (BA); –0.2275 (DPA), 0.0063 (HxA); 0.0553 (DBA), which point to the existence of interactions between unlike molecules and structural effects. On the other hand, along a given homologous series, $V_m^E$ increases with the amine size (Figures 1, 2). This means that, other than the phenomena which decrease $V_m^E$ (differences in size between components and lower positive contributions because of the disruption of amine-amine interactions), the predominant effects are: i) the higher number of broken interactions between DMA molecules by longer amines; and ii) the lower number and weaker DMA-amine interactions created in systems involving larger amines, as then the amine group is more sterically hindered. The replacement of HxA by DPA leads to a lower $V_m^E$ value, as in the case of the HxA or DPA + heptane mixtures (see above). Therefore, this trend can be explained by the decrease of the positive contribution to $V_m^E$ related to the breaking of interactions between like molecules when a secondary amine is involved. Interestingly, the same behavior is encountered in 1-alkanol + HxA or + DPA systems [51, 52]. The small positive $V_m^E$ values of the DBA solution over the whole concentration range underline the importance of the positive contribution to $V_m^E$ from the breaking of DMA-DMA interactions by the large aliphatic surface of DBA. In fact, the corresponding $V_m^E$ curve, skewed to higher $x_1$ values (Figure 1), reveals that DBA is a good breaker of the interactions between DMA molecules. The mentioned surface is smaller for HxA and then very small positive $V_m^E$ values are encountered at lower DMA concentrations (Figure 2). Negative $V_m^E$ values at the other side of the concentration range (Figure 2) suggest that interactions between unlike molecules are more favorable. This is consistent with the observed $V_m^E$ minimum of the BA mixture at $x_1 \approx 0.56$ (Figure 2). Interestingly, the symmetry of the $V_m^E$ curve of the DPA system is opposite to that of the BA solution (Figure 1). This feature together with $V_m^E$ (DMA + DPA) < $V_m^E$ (DMA + BA) (Table 3) suggest that structural effects become relevant in the system with DPA. Calculations using the PFP model are in agreement with this statement (see below).

The values of the derived properties $\kappa_S^E$, $\alpha_p^E$ are negative, while those of $c^E$ are positive (Figures 3-6, Table 6). In any case, all of them are rather small in absolute value, indicating that the studied systems show a nearly ideal behavior with respect to these properties. Nevertheless, it should be mentioned that negative values of $\kappa_S^E$, $\alpha_p^E$ and $A_p = \left(\Delta V_m^E / \Delta T\right)_p$ (–3·10$^{-3}$ cm$^3$·mol$^{-1}$·K$^{-1}$ (DPA); –2.8·10$^{-3}$ cm$^3$·mol$^{-1}$·K$^{-1}$ (BA)) are characteristic of systems where relevant interactions between unlike molecules and/or structural effects exist [19, 53]. On the other hand, the quantities $V_m^E$ and $\kappa_S^E$ change in line along a homologous series, while $c^E$ shows an opposite variation. The same behavior is observed when replacing HxA by DPA.

### 6.1. Internal pressures

The internal pressure [54-57], $P_{int}$, is an adequate quantity to examine the intermolecular forces in liquids and liquid mixtures:

$$P_{\text{int}} = T\frac{\alpha_p}{\kappa_T} - p \quad (17)$$

Here, the $\kappa_T$ values of the mixtures have been obtained from

$$\kappa_T = \kappa_S + \frac{TV_m(\alpha_p)^2}{C_{pm}} \quad (18)$$

assuming $C_{pm}^E = 0$ [58], and $\alpha_p^E = 0$ when experimental data are not available. For the pure compounds studied, $P_{\text{int}}^*$/MPa = 447.0 (DMA), 338.3 (BA), 343.5 (HxA), 303.4 (DPA), and 306.7 (DBA), whereas for the DMA mixtures $P_{\text{int}}$/MPa = 386.8 (BA), 381.9 (HxA), 353.2 (DPA), and 348.2 (DBA). The most important contributions to $P_{\text{int}}$ arise from dispersion forces and weak dipole-dipole interactions [56], and therefore these results suggest that dipolar interactions are stronger in the systems with linear primary amines.

We have also determined the excess internal pressures, $P_{\text{int}}^E = P_{\text{int}} - P_{\text{int}}^{id}$, ($P_{\text{int}}^{id} = T\alpha_p^{id}/\kappa_T^{id} - p$ [59]). Thus, $P_{\text{int}}^E$(DMA)/MPa = 10.8 (BA), 5.3 (HxA), 11.7 (DPA), and 6.3 (DBA). Systems with strong interactions between unlike molecules show large $P_{\text{int}}^E$ values. For example, $P_{\text{int}}^E$ = 61.4 MPa for the aniline + 2-propanone system [19]. This is seen to be verified by the above results, although the fact that the value for the HxA mixture is lower than for the DPA system may be due, at least partially, to structural effects, as they are similar to the hexane + hexadecane mixture (6.5 MPa [1, 60]). This is also consistent with the observed trend for their $V_m^E$ values [12-15, 17, 19].

The Van der Waals model allows to obtain the internal pressure from [55]:

$$P_{\text{int}}^{\text{VDW}} = \frac{RT}{x_1 V_{\text{fm},1}^* + x_2 V_{\text{fm},2}^* + V_m^E} - p \quad (19)$$

where $V_{\text{fm},i}^* = RT/(p + P_{\text{int},i}^*)$ is the free molar volume of the pure component $i$. The relative deviations of the results obtained from equation (19) and the experimental ones, $(P_{\text{int}}^{\text{VDW}} - P_{\text{int}})/P_{\text{int}}$, for the DMA mixtures are 2.7% (BA), 1.6% (HxA), 5.8% (DPA), and 3.6% (DBA). One can conclude that the Van der Waals equation is useful for the $P_{\text{int}}$ calculation of the studied solutions.

### 6.2. Molar refractions

The refractive index at optical wavelengths is closely related to dispersion forces, since the molar refraction (or molar refractivity), $R_m$, defined by the Lorentz-Lorenz equation [61, 62]:

$$R_m = \frac{n_D^2 - 1}{n_D^2 + 2} V_m = \frac{N_A \alpha_e}{3\varepsilon_0} \quad (20)$$

(where $N_A$ and $\varepsilon_0$ stand for Avogadro's constant and the vacuum permittivity, respectively) is proportional to the mean electronic contribution, $\alpha_e$, to the polarizability, [61]. The values of $R_m$ / cm$^3$·mol$^{-1}$ for the investigated systems are 24.7 (BA), 28.2 (HxA), 29.0 (DPA), and 33.6 (DBA). Clearly, dispersive interactions are more important for larger amines in a homologous series. Moreover, it can be stated that these forces are quite similar for the HxA and DPA systems, and therefore the corresponding difference in their $P_{int}$ values is principally due to dipolar interactions.

### 6.3. Comparison with other systems

For the considered amines, and also for aniline, mixtures with DMF are characterized by lower $V_m^E$ / cm$^3$·mol$^{-1}$ values: –0.2630 (BA), –0.0210 (HxA), –0.2893 (DPA), and 0.0178 (DBA) [20]; –0.6615 (aniline) [22], and –0.6092 cm$^3$·mol$^{-1}$ at 303.15 K for DMA + aniline [26]. This allows to conclude that amide-amine interactions are stronger in mixtures with DMF. Interestingly, deviations between experimental $P_{int}$ values and results from equation (12) are slightly larger for DMF systems: 7.6% (BA), 5.7% (HxA), 4.2% (DPA) and 3.1% (DBA) [20], which suggests that dipolar interactions are more relevant in such solutions. Finally, it is noteworthy that $V_m^E$ values are much lower for the mixture including aniline; this reveals that interactions between unlike molecules are strengthened when aniline is involved. The same trend is encountered for 2-alkanone + DPA or + aniline systems [12-19].

It is here pertinent to examine the effect of replacing a *N,N*-dialkylamide (DMF or DMA) by a 2-alkanone of similar size (2-propanone or 2-butanone). $V_m^E$ values of 2-propanone or 2-butanone + DPA, or + DBA mixtures are higher than those of the corresponding systems with DMF or DMA. For example, $V_m^E$(DPA)/cm$^3$·mol$^{-1}$= 0.243 (2-propanone) [12], 0.144 (2-butanone) [17] and $V_m^E$(DBA)/cm$^3$·mol$^{-1}$ = 0.417 (2-propanone) [12]; 0.265 (2-butanone) [17]. In addition, the $H_m^E$ values of these 2-alkanone mixtures are positive [16]. All this suggests that amide-amine interactions are stronger than alkanone-amine interactions in mixtures containing a linear secondary amine. Interestingly, aniline mixtures show a rather different behavior. For the 2-alkanone + aniline mixtures, we have $H_m^E$/ J·mol$^{-1}$ = –1236 (2-propanone); –1165 (2-butanone) [18] and $V_m^E$/cm$^3$·mol$^{-1}$ = –1.183 (2-propanone) [19]; –1.246 (2-butanone) [13]. The lower $V_m^E$ and the higher $H_m^E$ values of the 2-propanone mixture compared to those of the DMF system indicate that interactions between unlike molecules are stronger in the latter solution and that structural effects are more relevant in the 2-propanone system. Surprisingly, DMA-aniline interactions seem to be weaker than (2-butanone)-aniline interactions (see the corresponding $H_m^E$ values of these systems). This matter deserves a careful investigation, currently undertaken.

### 6.4. Prigogine-Flory-Patterson theory

In the framework of this theory, calculations show that $\chi_{12}$ increases when replacing DMF by DMA in systems with a given amine (component 2) (Table 8). In the original Flory model [37], $\chi_{12}$ is proportional to $\Delta\eta/v_s^*$, being $v_s^*$ the reduction volume of a segment and $\Delta\eta = \eta_{11} + \eta_{22} - 2\eta_{12}$. The positive $\eta_{ij}$ magnitudes characterize the energy of interaction for a pair of neighboring sites. As $\eta_{22}$ remains constant, the $\chi_{12}$ value increase may be due to the predominance of the $\eta_{12}$ decrease over that of $\eta_{11}$. The latter is linked to a weakening of the amide-amide interactions; the former merely reflects a weakening of the interactions between unlike molecules. Similar trends are also valid when BA is replaced by HxA in DMF solutions. Interestingly, the DMA + DPA or + DBA systems are characterized by the same $\chi_{12}$ value (Table 8), which suggests that such mixtures essentially differ in size effects. Finally, an inspection of the different contributions to $V_m^E$ listed in Table 8 shows that $(V_{m,\text{curvature}}^E + V_{m,p^*\text{effect}}^E)/V_m^E$ is much higher (in absolute value) for systems with DPA. This indicates that structural effects are more important for such a type of solution.

Regarding the composition dependence of $V_m^E$, the model describes fairly well this excess function for the systems DMF or DMA + BA, or + DPA (Figures 8,9). Results for the mixtures with DBA or HxA are somewhat poorer as the representation of low $V_m^E$ values is a very difficult task for any theoretical model.

## 7. Conclusions

Binary systems of DMA + BA, + HxA, + DPA or + DBA have been studied at different temperatures, reporting values of $\rho$, $c$, $n_D$ and of the excess functions ($V_m^E$, $\kappa_S^E$, $c^E$, $\alpha_p^E$ and $n_D^E$) determined from these ones. Negative and low positive $V_m^E$ values for the investigated mixtures point out to the existence of interactions between unlike molecules, as well as of structural effects. This is also supported by results from the PFP model. $V_m^E$ values are higher in the case of systems with linear primary amines, as the breaking of amine-amine interactions is more relevant than when linear secondary amines are involved. Steric hindrance of the amide group appears to be relevant, since comparisons with results of systems including amines and DMF or DMA show that interactions between unlike molecules are stronger in the former systems. The main differences between mixtures containing DPA and HxA come essentially from dipolar interactions. Dispersive interactions increase with the amine size in systems with a given amide.

## Acknowledgements

F. Hevia gratefully acknowledges the grant received from the program 'Ayudas para la Formación de Profesorado Universitario (convocatoria 2014), de los subprogramas de Formación y de Movilidad incluidos en el Programa Estatal de Promoción del Talento y su Empleabilidad, en el marco del Plan Estatal de Investigación Científica y Técnica y de Innovación 2013-2016, de la Secretaría de Estado de Educación, Formación Profesional y Universidades, Ministerio de Educación, Cultura y Deporte, Gobierno de España'.

Table 1

Sample description

| Chemical | CAS number | Source | Method of purification | Purity[a] | Analysis method |
|---|---|---|---|---|---|
| *N,N*-dimethylacetamide (DMA) | 127-19-5 | Sigma-Aldrich | None | ≥ 0.995 | GC[b] |
| *N*-propylpropan-1-amine (DPA) | 142-84-7 | Aldrich | None | ≥ 0.99 | GC[b] |
| *N*-butylbutan-1-amine (DBA) | 111-92-2 | Aldrich | None | ≥ 0.995 | GC[b] |
| butan-1-amine (BA) | 109-73-9 | Sigma-Aldrich | None | ≥ 0.995 | GC[b] |
| hexan-1-amine (HxA) | 111-26-2 | Aldrich | None | ≥ 0.995 | GC[b] |

[a] In mole fraction.

[b] Gas chromatography.

Table 2

Physical properties of pure compounds at temperature $T$ and pressure $p = 0.1$ MPa. [a]

| Property | $T$/K | DMF | DPA | DBA | BA | HxA |
|---|---|---|---|---|---|---|
| $\rho^*$/g·cm$^{-3}$ | 293.15 | 0.94087 | 0.73778 | 0.75970 | 0.73705 | 0.76439 |
| | | 0.940846 [63] | 0.7375 [1] | 0.759571 [17] | 0.73712 [64] | 0.7651 [65] |
| | 298.15 | 0.93630 | 0.73322 | 0.75553 | 0.73233 | 0.76019 |
| | | 0.936233 [63] | 0.73321 [66] | 0.755457 [17] | 0.73233 [64] | 0.76013 [67] |
| | 303.15 | 0.93169 | 0.72870 | 0.75146 | 0.72750 | 0.75589 |
| | | 0.931618 [63] | 0.729087 [17] | 0.751329 [17] | 0.72751 [64] | 0.7562 [65] |
| $c^*$/m·s$^{-1}$ | 293.15 | 1475.1 | 1208.7 | 1261.1 | 1268.1 | 1323.9 |
| | | | 1209 [17] | 1261.2 [17] | | |
| | 298.15 | 1455.7 | 1187.3 | 1241.5 | 1246.1 | 1303.8 |
| | | 1455.37 [68] | 1198 [70] | 1248 [70] | 1247.8 [71] | 1304.7 [67] |
| | | 1458 [69] | | | | |
| | 303.15 | 1435.7 | 1166.7 | 1222.5 | 1224.5 | 1283.5 |
| | | 1441 [72] | 1174 [70] | 1227 [70] | 1227 [70] | 1285 [70] |
| $\alpha_p^*$/10$^{-3}$K$^{-1}$ | 298.15 | 0.980 | 1.239 | 1.090 | 1.304 | 1.119 |
| | | 0.960 [73] | 1.29 [1] | 1.12 [1] | 1.314 [70] | 1.13 [67] |
| $\kappa_S^*$/TPa$^{-1}$ | 293.15 | 488.5 | 927.8 | 827.7 | 843.7 | 746.4 |
| | | | 926.5 [70] | | | |
| | 298.15 | 504.0 | 967.5 | 858.7 | 879.4 | 773.9 |
| | | 504.29 [68] | 947 [70] | 849 [70] | 876.6 [71] | 773 [67] |
| | 303.15 | 520.7 | 1008.2 | 890.4 | 916.7 | 803.1 |
| | | 516 [72] | 992 [70] | 883 [70] | 912 [70] | 800 [70] |
| $\kappa_T^*$/TPa$^{-1}$ | 298.15 | 653.5 | 1217.3 | 1059.4 | 1148.7 | 971.1 |
| | | 671 [74] | 1183 [70] | 1039 [70] | 1145 [70] | 975 [67] |
| $C_{pm}^*$/J·mol$^{-1}$·K$^{-1}$ | 298.15 | 178.2 [75] | 252.84 [1] | 302 [70] | 188 [76] | 252 [76] |
| $n_D^*$ | 293.15 | 1.43814 | 1.40398 | | 1.40059 | |
| | | 1.4384 [1] | 1.4043 [1] | | | |
| | 298.15 | 1.43595 | 1.40135 | 1.41488 | 1.39789 | 1.41571 |
| | | 1.4363 [69] | 1.40132 [77] | 1.4152 [1] | 1.3987 [1] | 1.4160 [70] |
| | 303.15 | 1.43382 | 1.39871 | | 1.39507 | |
| | | 1.4342 [69] | 1.4022 [70] | | 1.3978 [70] | |

[a] $\rho^*$, density; $c^*$, speed of sound; $\alpha_p^*$, isobaric thermal expansion coefficient; $\kappa_S^*$, adiabatic compressibility; $\kappa_T^*$, isothermal compressibility; $C_{pm}^*$, isobaric molar heat capacity; and $n_D^*$, refractive index. The standard uncertainties are: $u(T) = 0.01$ K (for $n_D^*$ values, $u(T) = 0.02$ K); $u(p) = 1$ kPa; $u(c^*) = 0.2$ m·s$^{-1}$; $u(\rho^*) = 0.00005$ g·cm$^{-3}$; $u(n_D^*) = 0.00008$ and (relative values) $u_r(\alpha_p^*) = 0.015$; $u_r(\kappa_S^*) = 0.002$; $u_r(\kappa_T^*) = 0.012$.

Table 3

Densities, $\rho$, excess molar volumes, $V_m^E$, and speeds of sound, $c$, for N,N-dimethylacetamide (1) + amine (2) mixtures at temperature $T$ and pressure $p = 0.1$ MPa. [a]

| $x_1$ | $\rho$/g·cm$^{-3}$ | $V_m^E$/cm$^3$·mol$^{-1}$ | $c$/m·s$^{-1}$ | $x_1$ | $\rho$/g·cm$^{-3}$ | $V_m^E$/cm$^3$·mol$^{-1}$ | $c$/m·s$^{-1}$ |
|---|---|---|---|---|---|---|---|
| \multicolumn{8}{c}{DMA (1) + DPA (2); $T$/K= 293.15} | | | | | | | |
| 0.0000 | 0.73778 |  | 1208.7 | 0.4914 | 0.81947 | –0.2137 | 1304.1 |
| 0.0621 | 0.74684 | –0.0656 | 1218.8 | 0.5582 | 0.83282 | –0.2096 | 1321.1 |
| 0.1201 | 0.75556 | –0.1122 | 1228.6 | 0.6520 | 0.85273 | –0.1936 | 1347.3 |
| 0.1432 | 0.75911 | –0.1272 | 1232.6 | 0.7141 | 0.86668 | –0.1719 | 1366.3 |
| 0.2142 | 0.77035 | –0.1659 | 1245.4 | 0.7604 | 0.87751 | –0.1495 | 1381.5 |
| 0.2434 | 0.77511 | –0.1780 | 1250.8 | 0.8012 | 0.88743 | –0.1312 | 1395.5 |
| 0.3154 | 0.78722 | –0.1971 | 1264.9 | 0.8494 | 0.89962 | –0.1098 | 1413.2 |
| 0.3398 | 0.79148 | –0.2052 | 1269.9 | 0.9017 | 0.91334 | –0.0754 | 1433.4 |
| 0.4140 | 0.80484 | –0.2178 | 1286.0 | 0.9457 | 0.92538 | –0.0441 | 1451.5 |
| 0.4668 | 0.81475 | –0.2175 | 1298.2 | 1.0000 | 0.94087 |  | 1475.1 |
| \multicolumn{8}{c}{DMA (1) + DPA (2); $T$/K= 298.15} | | | | | | | |
| 0.0000 | 0.73322 |  | 1187.3 | 0.5678 | 0.83024 | –0.2214 | 1303.8 |
| 0.0668 | 0.74299 | –0.0767 | 1198.4 | 0.5999 | 0.83691 | –0.2120 | 1312.5 |
| 0.1010 | 0.74814 | –0.1103 | 1204.2 | 0.6543 | 0.84864 | –0.1998 | 1328.2 |
| 0.1466 | 0.75510 | –0.1400 | 1212.2 | 0.7153 | 0.86240 | –0.1823 | 1347.1 |
| 0.2032 | 0.76403 | –0.1719 | 1222.5 | 0.7605 | 0.87301 | –0.1619 | 1362.0 |
| 0.2606 | 0.77342 | –0.1967 | 1233.5 | 0.8006 | 0.88277 | –0.1440 | 1375.9 |
| 0.3112 | 0.78198 | –0.2125 | 1243.6 | 0.8576 | 0.89720 | –0.1137 | 1396.8 |
| 0.3584 | 0.79022 | –0.2214 | 1253.4 | 0.8960 | 0.90729 | –0.0875 | 1411.8 |
| 0.3933 | 0.79648 | –0.2258 | 1261.1 | 0.9495 | 0.92191 | –0.0469 | 1433.8 |
| 0.4622 | 0.80932 | –0.2314 | 1277.0 | 1.0000 | 0.93630 |  | 1455.7 |
| 0.5019 | 0.81700 | –0.2303 | 1286.7 |  |  |  |  |
| \multicolumn{8}{c}{DMA (1) + DPA (2); $T$/K= 303.15} | | | | | | | |
| 0.0000 | 0.72870 |  | 1166.7 | 0.5632 | 0.82479 | –0.2412 | 1282.9 |
| 0.0609 | 0.73755 | –0.0658 | 1176.8 | 0.5926 | 0.83089 | –0.2360 | 1290.9 |
| 0.1008 | 0.74351 | –0.1011 | 1183.7 | 0.6511 | 0.84343 | –0.2204 | 1307.7 |
| 0.1975 | 0.75853 | –0.1674 | 1201.1 | 0.7089 | 0.85641 | –0.2019 | 1325.4 |
| 0.2416 | 0.76569 | –0.1905 | 1209.6 | 0.7618 | 0.86882 | –0.1802 | 1342.8 |
| 0.2915 | 0.77407 | –0.2173 | 1219.5 | 0.7881 | 0.87517 | –0.1655 | 1351.9 |
| 0.3409 | 0.78258 | –0.2288 | 1229.8 | 0.8571 | 0.89250 | –0.1226 | 1376.9 |
| 0.3957 | 0.79240 | –0.2427 | 1241.8 | 0.9026 | 0.90448 | –0.0903 | 1394.6 |
| 0.4536 | 0.80316 | –0.2476 | 1255.2 | 0.9464 | 0.91644 | –0.0520 | 1412.6 |
| 0.4910 | 0.81033 | –0.2451 | 1264.2 | 1.0000 | 0.93169 | | 1435.7 |

DMA (1) + DBA (2); $T$/K= 298.15

| | | | | | | | |
|---|---|---|---|---|---|---|---|
| 0.0000 | 0.75553 | | 1241.5 | 0.5040 | 0.81954 | 0.0540 | 1304.4 |
| 0.0556 | 0.76111 | 0.0057 | 1246.5 | 0.6059 | 0.83747 | 0.0568 | 1324.5 |
| 0.1101 | 0.76685 | 0.0150 | 1251.6 | 0.6466 | 0.84529 | 0.0572 | 1333.6 |
| 0.1416 | 0.77031 | 0.0202 | 1254.8 | 0.6971 | 0.85562 | 0.0542 | 1346.1 |
| 0.2017 | 0.77724 | 0.0263 | 1261.3 | 0.7538 | 0.86809 | 0.0502 | 1361.6 |
| 0.2645 | 0.78493 | 0.0327 | 1268.7 | 0.7900 | 0.87660 | 0.0437 | 1372.5 |
| 0.3006 | 0.78956 | 0.0411 | 1273.2 | 0.8612 | 0.89469 | 0.0332 | 1396.6 |
| 0.3448 | 0.79551 | 0.0436 | 1279.2 | 0.8925 | 0.90328 | 0.0271 | 1408.3 |
| 0.3993 | 0.80324 | 0.0513 | 1287.1 | 0.9566 | 0.92228 | 0.0106 | 1435.3 |
| 0.4461 | 0.81028 | 0.0526 | 1294.4 | 1.0000 | 0.93629 | | 1455.7 |

DMA (1) + BA (2); $T$/K= 293.15

| | | | | | | | |
|---|---|---|---|---|---|---|---|
| 0.0000 | 0.73705 | | 1268.1 | 0.5968 | 0.85698 | –0.1771 | 1383.5 |
| 0.0490 | 0.74668 | –0.0355 | 1276.6 | 0.6577 | 0.86951 | –0.1624 | 1396.8 |
| 0.1067 | 0.75803 | –0.0680 | 1286.8 | 0.7542 | 0.88954 | –0.1354 | 1418.5 |
| 0.1477 | 0.76612 | –0.0867 | 1294.1 | 0.8499 | 0.90960 | –0.1022 | 1440.4 |
| 0.2508 | 0.78662 | –0.1253 | 1313.2 | 0.9063 | 0.92143 | –0.0716 | 1453.5 |
| 0.3494 | 0.80649 | –0.1611 | 1332.2 | 0.9420 | 0.92898 | –0.0526 | 1461.8 |
| 0.4513 | 0.82709 | –0.1687 | 1352.7 | 1.0000 | 0.94108 | | 1475.2 |
| 0.5505 | 0.84739 | –0.1736 | 1373.5 | | | | |

DMA (1) + BA (2); $T$/K= 298.15

| | | | | | | | |
|---|---|---|---|---|---|---|---|
| 0.0000 | 0.73233 | | 1246.1 | 0.5646 | 0.84566 | –0.1948 | 1356.4 |
| 0.0536 | 0.74283 | –0.0378 | 1255.7 | 0.6946 | 0.87259 | –0.1823 | 1385.3 |
| 0.1214 | 0.75625 | –0.0862 | 1268.1 | 0.7540 | 0.88492 | –0.1607 | 1398.7 |
| 0.1929 | 0.77041 | –0.1198 | 1281.2 | 0.8534 | 0.90573 | –0.1210 | 1421.8 |
| 0.2540 | 0.78258 | –0.1418 | 1292.8 | 0.9055 | 0.91672 | –0.0960 | 1434.2 |
| 0.3630 | 0.80457 | –0.1808 | 1314.3 | 0.9446 | 0.92485 | –0.0606 | 1443.0 |
| 0.4684 | 0.82594 | –0.1925 | 1335.9 | 1.0000 | 0.93633 | | 1455.6 |
| 0.5051 | 0.83343 | –0.1938 | 1343.6 | | | | |

DMA (1) + BA (2); $T$/K= 303.15

| | | | | | | | |
|---|---|---|---|---|---|---|---|
| 0.0000 | 0.72750 | | 1224.5 | 0.4971 | 0.82713 | –0.2009 | 1321.7 |
| 0.0517 | 0.73770 | –0.0456 | 1233.9 | 0.6574 | 0.86025 | –0.1930 | 1356.9 |
| 0.1082 | 0.74889 | –0.0875 | 1244.3 | 0.7957 | 0.88919 | –0.1582 | 1388.8 |
| 0.1535 | 0.75783 | –0.1083 | 1252.8 | 0.8498 | 0.90040 | –0.1161 | 1401.1 |
| 0.2556 | 0.77826 | –0.1617 | 1272.3 | 0.9090 | 0.91297 | –0.0918 | 1415.1 |
| 0.2921 | 0.78562 | –0.1783 | 1279.6 | 0.9404 | 0.91948 | –0.0590 | 1422.3 |
| 0.3582 | 0.79885 | –0.1832 | 1292.6 | 1.0000 | 0.93195 | | 1436.1 |
| 0.4583 | 0.81922 | –0.2014 | 1313.3 | | | | |

DMA (1) + HxA (2); $T$/K= 298.15

| | | | | | | | |
|---|---|---|---|---|---|---|---|
| 0.0000 | 0.76019 | | 1303.8 | 0.6130 | 0.85274 | 0.0027 | 1374.5 |
| 0.0575 | 0.76737 | 0.0052 | 1308.5 | 0.6994 | 0.86931 | –0.0020 | 1389.5 |
| 0.1197 | 0.77545 | 0.0052 | 1314.0 | 0.7620 | 0.88202 | –0.0064 | 1401.2 |
| 0.1582 | 0.78060 | 0.0084 | 1317.5 | 0.8060 | 0.89133 | –0.0094 | 1410.3 |
| 0.2079 | 0.78743 | 0.0114 | 1322.4 | 0.8544 | 0.90192 | –0.0086 | 1420.5 |
| 0.2450 | 0.79271 | 0.0087 | 1326.1 | 0.8959 | 0.91134 | –0.0090 | 1430.2 |
| 0.3068 | 0.80176 | 0.0082 | 1333.0 | 0.9467 | 0.92330 | –0.0066 | 1442.2 |
| 0.4037 | 0.81674 | 0.0063 | 1344.4 | 1.0000 | 0.93637 | | 1455.8 |
| 0.5122 | 0.83471 | 0.0068 | 1359.1 | | | | |

[a] The standard uncertainties are: $u(x_1) = 0.0001$; $u(p) = 1\,\text{kPa}$; $u(T) = 0.01\,\text{K}$. The standard uncertainties are: $u(\rho) = 0.00005$ g·cm$^{-3}$; $u(V_m^E) = (0.010\,|V_{m,max}^E| + 0.005\,\text{cm}^3\cdot\text{mol}^{-1})$; $u(c) = 0.2$ m·s$^{-1}$.

Table 4

Excess functions, at temperature $T = 298.15$ K and pressure $p = 0.1$ MPa, for $\kappa_S$, adiabatic compressibility, $c$, speed of sound, and $\alpha_p$, isobaric thermal expansion coefficient, of N,N-dimethylacetamide (1) + amine (2) mixtures. [a]

| $x_1$ | $\kappa_S^E$/TPa$^{-1}$ | $c^E$/m·s$^{-1}$ | $\alpha_p^E$/10$^{-6}$·K$^{-1}$ [b] | $x_1$ | $\kappa_S^E$/TPa$^{-1}$ | $c^E$/m·s$^{-1}$ | $\alpha_p^E$/10$^{-6}$·K$^{-1}$ [b] |
|---|---|---|---|---|---|---|---|
| DMA (1) + DPA (2) | | | | | | | |
| 0.0668 | −9.7 | 5.8 | −1 | 0.5678 | −45.0 | 38.3 | −26 |
| 0.1010 | −14.1 | 8.6 | −2 | 0.5999 | −44.6 | 39.1 | −26 |
| 0.1466 | −19.5 | 12.3 | −4 | 0.6543 | −43.4 | 39.9 | −26 |
| 0.2032 | −25.6 | 16.7 | −7 | 0.7153 | −40.5 | 39.7 | −26 |
| 0.2606 | −31.1 | 21.0 | −10 | 0.7605 | −37.3 | 38.3 | −24 |
| 0.3112 | −35.1 | 24.5 | −14 | 0.8006 | −33.6 | 36.1 | −22 |
| 0.3584 | −38.2 | 27.6 | −16 | 0.8576 | −26.7 | 30.7 | −18 |
| 0.3933 | −40.3 | 29.8 | −19 | 0.8960 | −20.9 | 25.4 | −14 |
| 0.4622 | −43.2 | 33.7 | −22 | 0.9495 | −11.2 | 14.7 | −7 |
| 0.5019 | −44.3 | 35.7 | −24 | | | | |
| DMA (1) + DBA (2) | | | | | | | |
| 0.0556 | −2.3 | 1.7 | | 0.5040 | −16.7 | 15.2 | |
| 0.1101 | −4.3 | 3.3 | | 0.6059 | −18.1 | 17.5 | |
| 0.1416 | −5.5 | 4.3 | | 0.6466 | −18.2 | 18.1 | |
| 0.2017 | −7.8 | 6.2 | | 0.6971 | −18.1 | 18.8 | |
| 0.2645 | −10.1 | 8.1 | | 0.7538 | −17.2 | 18.7 | |
| 0.3006 | −11.2 | 9.2 | | 0.7900 | −16.2 | 18.3 | |
| 0.3448 | −12.7 | 10.6 | | 0.8612 | −13.1 | 15.9 | |
| 0.3993 | −14.3 | 12.3 | | 0.8925 | −11.0 | 13.9 | |
| 0.4461 | −15.4 | 13.6 | | 0.9566 | −5.4 | 7.4 | |
| DMA (1) + BA (2) | | | | | | | |
| 0.0536 | −7.7 | 5.4 | −11 | 0.5646 | −35.9 | 35.0 | −31 |
| 0.1214 | −16.2 | 11.7 | −20 | 0.6946 | −31.3 | 33.6 | −31 |
| 0.1929 | −23.1 | 17.5 | −25 | 0.7540 | −27.4 | 30.8 | −31 |
| 0.2540 | −27.9 | 22.0 | −27 | 0.8534 | −18.8 | 22.9 | −26 |
| 0.3630 | −33.9 | 28.7 | −29 | 0.9055 | −13.2 | 16.8 | −22 |
| 0.4684 | −36.4 | 33.1 | −30 | 0.9446 | −8.1 | 10.6 | −16 |
| 0.5051 | −36.5 | 34.1 | −30 | | | | |
| DMA (1) + HxA (2); $T$/K= 298.15 | | | | | | | |
| 0.0575 | −1.8 | 1.6 | | 0.6130 | −12.2 | 13.3 | |
| 0.1197 | −3.8 | 3.4 | | 0.6994 | −11.7 | 13.5 | |

| | | | | | | |
|---|---|---|---|---|---|---|
| 0.1582 | –4.9 | 4.3 | | 0.7620 | –10.6 | 12.6 |
| 0.2079 | –6.3 | 5.7 | | 0.8060 | –9.7 | 11.9 |
| 0.2450 | –7.1 | 6.6 | | 0.8544 | –7.9 | 10.0 |
| 0.3068 | –8.8 | 8.3 | | 0.8959 | –6.4 | 8.3 |
| 0.4037 | –10.5 | 10.3 | | 0.9467 | –3.5 | 4.8 |
| 0.5122 | –11.9 | 12.3 | | | | |

[a] The standard uncertainties are: $u(x_1) = 0.0001$; $u(p) = 1\,\text{kPa}$; $u(T) = 0.01\,\text{K}$. The standard uncertainties are: $u(c^\text{E}) = 0.4$; and (relative values) $u_\text{r}(\kappa_S^\text{E}) = 0.015$; $u_\text{r}(\alpha_p^\text{E}) = 0.025$.

[b] Density values at 293.15 and 303.15 K at the mole fractions reported at 298.15 K were obtained from the corresponding Redlich-Kister adjustments for $V_\text{m}^\text{E}$.

Table 5

Refractive indices, $n_D$, and the corresponding excess values, $n_D^E$, of N,N-dimethylacetamide (1) + amine (2) mixtures at temperature $T$ and pressure $p = 0.1$ MPa. [a]

| $x_1$ | $n_D$ | $n_D^E/10^{-5}$ | $x_1$ | $n_D$ | $n_D^E/10^{-5}$ |
|---|---|---|---|---|---|
| \multicolumn{6}{c}{DMA (1) + DPA (2); $T/K = 293.15$} ||||||
| 0.0000 | 1.40398 |    | 0.5582 | 1.42085 | 104 |
| 0.0621 | 1.40569 | 23 | 0.6520 | 1.42418 | 102 |
| 0.1201 | 1.40731 | 42 | 0.7141 | 1.42649 | 97  |
| 0.1432 | 1.40797 | 49 | 0.7604 | 1.42826 | 90  |
| 0.2142 | 1.41003 | 69 | 0.8012 | 1.42986 | 82  |
| 0.3154 | 1.41303 | 87 | 0.8494 | 1.43179 | 69  |
| 0.3398 | 1.41376 | 89 | 0.9017 | 1.43396 | 52  |
| 0.4140 | 1.41609 | 99 | 0.9457 | 1.43581 | 32  |
| 0.4668 | 1.41780 | 103 | 1.0000 | 1.43814 |    |
| 0.4914 | 1.41860 | 104 |   |   |   |
| \multicolumn{6}{c}{DMA (1) + DPA (2); $T/K = 298.15$} ||||||
| 0.0000 | 1.40135 |    | 0.4622 | 1.41524 | 110 |
| 0.0668 | 1.40324 | 28 | 0.5678 | 1.41882 | 111 |
| 0.1010 | 1.40422 | 41 | 0.6543 | 1.42192 | 107 |
| 0.1466 | 1.40554 | 56 | 0.7153 | 1.42421 | 101 |
| 0.2032 | 1.40720 | 72 | 0.7605 | 1.42597 | 94  |
| 0.2606 | 1.40892 | 86 | 0.8006 | 1.42755 | 85  |
| 0.3112 | 1.41045 | 95 | 0.8576 | 1.42988 | 70  |
| 0.3584 | 1.41191 | 101 | 0.9495 | 1.43376 | 31 |
| 0.3933 | 1.41302 | 106 | 1.0000 | 1.43595 |    |
| \multicolumn{6}{c}{DMA (1) + DPA (2); $T/K = 303.15$} ||||||
| 0.0000 | 1.39871 |    | 0.5632 | 1.41619 | 105 |
| 0.0609 | 1.40041 | 21 | 0.5926 | 1.41723 | 104 |
| 0.1008 | 1.40155 | 35 | 0.6511 | 1.41934 | 97  |
| 0.1975 | 1.40440 | 65 | 0.7089 | 1.42153 | 91  |
| 0.2416 | 1.40577 | 79 | 0.7881 | 1.42468 | 79  |
| 0.2915 | 1.40727 | 87 | 0.8571 | 1.42755 | 63  |
| 0.3409 | 1.40885 | 99 | 0.9026 | 1.42948 | 46  |
| 0.3957 | 1.41059 | 104 | 0.9464 | 1.43142 | 29 |
| 0.4536 | 1.41246 | 106 | 1.0000 | 1.43382 |    |
| 0.4910 | 1.41370 | 106 |   |   |   |
| \multicolumn{6}{c}{DMA (1) + DBA (2); $T/K = 298.15$} ||||||
| 0.0000 | 1.41495 |    | 0.5543 | 1.42341 |    |
| 0.0554 | 1.41559 |    | 0.6061 | 1.42456 |    |
| 0.1101 | 1.41627 | | 0.6466 | 1.42545 | |
| 0.1413 | 1.41667 | | 0.6971 | 1.42667 | |
| 0.2106 | 1.41761 | | 0.7538 | 1.42813 | |
| 0.2645 | 1.41838 | | 0.7900 | 1.42907 | |
| 0.3006 | 1.41893 | | 0.8613 | 1.43120 | |
| 0.3448 | 1.41961 | | 0.8925 | 1.43216 | |
| 0.3993 | 1.42053 | | 0.9566 | 1.43436 | |
| 0.4461 | 1.42137 | | 1.0000 | 1.43592 | |
| 0.5040 | 1.42243 | | | | |
| | | DMA (1) + BA (2); $T/\text{K} = 293.15$ | | | |
| 0.0000 | 1.40059 | | 0.5505 | 1.42171 | 98 |
| 0.0490 | 1.40249 | 16 | 0.5968 | 1.42344 | 96 |
| 0.1067 | 1.40474 | 34 | 0.6577 | 1.42571 | 91 |
| 0.1477 | 1.40633 | 46 | 0.6982 | 1.42722 | 87 |
| 0.1916 | 1.40804 | 58 | 0.7542 | 1.42931 | 80 |
| 0.2508 | 1.41031 | 70 | 0.8499 | 1.43280 | 58 |
| 0.2962 | 1.41207 | 79 | 0.9063 | 1.43483 | 40 |
| 0.3494 | 1.41411 | 88 | 0.9420 | 1.43611 | 28 |
| 0.4513 | 1.41796 | 95 | 1.0000 | 1.43813 | |
| 0.5007 | 1.41983 | 97 | | | |
| | | DMA (1) + BA (2); $T/\text{K} = 298.15$ | | | |
| 0.0000 | 1.39789 | | 0.5979 | 1.42105 | 93 |
| 0.0536 | 1.39999 | 17 | 0.6554 | 1.42323 | 89 |
| 0.1214 | 1.40265 | 37 | 0.6946 | 1.42471 | 85 |
| 0.1929 | 1.40547 | 57 | 0.7540 | 1.42698 | 80 |
| 0.2540 | 1.40786 | 71 | 0.7912 | 1.42837 | 73 |
| 0.2974 | 1.40955 | 79 | 0.8534 | 1.43067 | 58 |
| 0.4027 | 1.41359 | 90 | 0.9055 | 1.43259 | 43 |
| 0.4684 | 1.41610 | 93 | 0.9446 | 1.43399 | 27 |
| 0.5051 | 1.41751 | 94 | 1.0000 | 1.43595 | |
| 0.5646 | 1.41978 | 94 | | | |
| | | DMA (1) + BA (2); $T/\text{K} = 303.15$ | | | |
| 0.0000 | 1.39507 | | 0.5583 | 1.41717 | 107 |
| 0.0517 | 1.39714 | 18 | 0.5962 | 1.41864 | 106 |
| 0.1082 | 1.39943 | 39 | 0.6574 | 1.42099 | 101 |
| 0.1535 | 1.40125 | 53 | 0.6970 | 1.42250 | 96 |
| 0.1894 | 1.40271 | 66 | 0.7537 | 1.42467 | 88 |
| 0.2556 | 1.40535 | 82 | 0.7957 | 1.42629 | 82 |
| 0.2921 | 1.40678 | 88 | 0.8498 | 1.42834 | 70 |
| 0.3582 | 1.40941 | 100 | 0.9090 | 1.43054 | 50 |
| 0.4583 | 1.41329 | 106 | 0.9404 | 1.43168 | 37 |

| | | | | |
|---|---|---|---|---|
| 0.4971 | 1.41479 | 106 | 1.0000 | 1.43375 |
| | | DMA (1) + HxA (2); $T$/K = 298.15 | | |
| 0.0000 | 1.41571 | | 0.6130 | 1.42641 |
| 0.0575 | 1.41652 | | 0.6994 | 1.42832 |
| 0.1197 | 1.41744 | | 0.7620 | 1.42978 |
| 0.1582 | 1.41804 | | 0.8060 | 1.43085 |
| 0.2079 | 1.41882 | | 0.8544 | 1.43205 |
| 0.2450 | 1.41942 | | 0.8959 | 1.43310 |
| 0.3068 | 1.42049 | | 0.9467 | 1.43445 |
| 0.4037 | 1.42224 | | 1.0000 | 1.43590 |
| 0.5122 | 1.42433 | | | |

[a] The standard uncertainties are: $u(x_1) = 0.0001$; $u(T) = 0.02$ K; $u(p) = 1$ kPa.; $u(n_D) = 0.00008$; $u(n_D^E) = 0.0002$.

Table 6

Coefficients $A_i$ and standard deviations, $\sigma(F^E)$ (equation (9)), for the representation of the $F^E$ property at temperature $T$ and pressure $p = 0.1$ MPa for $N,N$-dimethylacetamide (1) + amine (2) systems by equation (8).

| System | T/K | Property [a] $F^E$ | $A_0$ | $A_1$ | $A_2$ | $A_3$ | $A_4$ | $\sigma(F^E)$ |
|---|---|---|---|---|---|---|---|---|
| DMA + DPA | 293.15 | $V_m^E$ | −0.861 | 0.132 | −0.15 | | | 0.002 |
| | | $n_D^E/10^{-5}$ | 416 | 69 | 118 | 69 | | 0.7 |
| | 298.15 | $V_m^E$ | −0.910 | 0.140 | −0.23 | | | 0.0018 |
| | | $\kappa_S^E$ | −176.9 | −43.1 | −20.5 | | | 0.07 |
| | | $c^E$ | 142.3 | 90.7 | 54 | 33 | 20 | 0.06 |
| | | $\alpha_p^E/10^{-6}$ | −94.8 | −79 | 14 | | | 0.4 |
| | | $n_D^E/10^{-5}$ | 445.5 | 43 | 127 | 85 | | 0.4 |
| | 303.15 | $V_m^E$ | −0.983 | 0.063 | −0.12 | | | 0.0017 |
| | | $n_D^E/10^{-5}$ | 428 | −17 | 50 | 176 | | 1.1 |
| DMA + DBA | 298.15 | $V_m^E$ | 0.222 | 0.08 | | | | 0.0015 |
| | | $\kappa_S^E$ | −66.4 | −38.7 | −21.1 | −11 | | 0.06 |
| | | $c^E$ | 60.4 | 51.4 | 35 | 32 | 21 | 0.06 |
| DMA + BA | 293.15 | $V_m^E$ | −0.696 | −0.07 | −0.11 | | | 0.004 |
| | | $n_D^E/10^{-5}$ | 390 | 32 | 44 | 67 | | 0.7 |
| | 298.15 | $V_m^E$ | −0.77 | −0.14 | −0.21 | | | 0.005 |
| | | $\kappa_S^E$ | −145.9 | −0.6 | −9.0 | | | 0.12 |
| | | $c^E$ | 135.7 | 48 | 23 | 9 | | 0.13 |
| | | $\alpha_p^E/10^{-6}$ | −119 | −25 | −156 | | | 1 |
| | | $n_D^E/10^{-5}$ | 378 | 23 | 74 | 110 | | 0.8 |
| | 303.15 | $V_m^E$ | −0.81 | −0.07 | −0.24 | | | 0.005 |
| | | $n_D^E/10^{-5}$ | 429 | 3 | 105 | 189 | | 1 |
| DMA + HxA | 298.15 | $V_m^E$ | 0.024 | −0.06 | −0.07 | −0.08 | | 0.0014 |
| | | $\kappa_S^E$ | −47.0 | −19.3 | −6.7 | | | 0.1 |
| | | $c^E$ | 48.3 | 30.7 | 17.3 | 8 | | 0.1 |

[a] $F^E = V_m^E$, units: cm³·mol⁻¹; $F^E = c^E$, units: m·s⁻¹; $F^E = \kappa_S^E$ units: TPa⁻¹; $F^E = \alpha_p^E/10^{-6}$, units: K⁻¹.

Table 7

Values of molar volume, $V_{m,i}$, and Flory reduction parameters for volume, $V_{m,i}^*$, and pressure, $p_i^*$, at 298.15 K of pure compounds.

| Compound | $V_{m,i}$ /cm$^3$·mol$^{-1}$ | $V_{m,i}^*$ / cm$^3$· mol$^{-1}$ | $p_i^*$ /J·cm$^{-3}$ |
|---|---|---|---|
| DMF[b] | 77.42 | 61.97 | 711.4 |
| DMA | 93.05 | 74.82 | 691.4 |
| DPA | 138.00 | 106.59 | 508.8 |
| DBA | 171.13 | 135.82 | 491.7 |
| BA | 99.87 | 76.42 | 578.0 |
| HxA | 133.11 | 104.68 | 555.6 |

[a] Values determined using densities, thermal expansion coefficients and isothermal compressibilities given in Table 2 and in reference [20].

[b] *N,N*-dimethylformamide.

Table 8

Values of the contributions to $V_m^E$, at 298.15 K and equimolar composition, for DMF or DMA + amine systems calculated according to the PFP model (equations (10)-(12)) using the interaction parameters, $\chi_{12}$, also listed.

| System[a] | $\chi_{12}^b$ /J·cm$^{-3}$ | $V_m^E$ contributions / cm$^3$·mol$^{-1}$ | | |
|---|---|---|---|---|
| | | Interactional term | Curvature term | $p^*$ effect term |
| DMF + BA | 2.5 | 0.0284 | –0.085 | –0.206 |
| DMF + HxA | 8.4 | 0.1001 | –0.015 | –0.110 |
| DMF + DPA | 6.2 | 0.0837 | –0.064 | –0.308 |
| DMF + DBA | 8.15 | 0.1110 | –0.004 | –0.089 |
| DMA + BA | 9.3 | 0.1180 | –0.111 | –0.214 |
| DMA + HxA | 12.42 | 0.1680 | –0.026 | –0.135 |
| DMA + DPA | 13.45 | 0.2035 | –0.088 | –0.351 |
| DMA + DBA | 13.45 | 0.2069 | –0.010 | –0.140 |

[a] DMF, *N,N*-dimethylformamide.

[b] Determined from $V_m^E$ data at equimolar composition [20], this work.

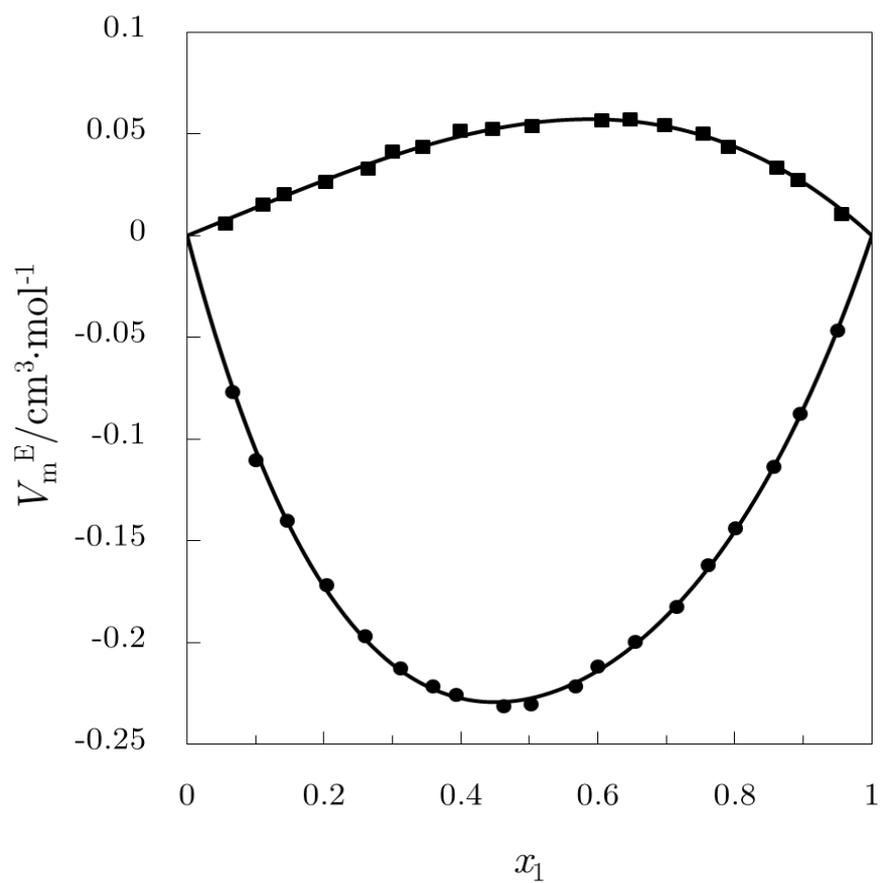

Figure 1

Excess molar volumes, $V_m^E$, for DMA (1) + DPA (2), or + DBA (2) systems at 0.1 MPa and 298.15 K. Full symbols, experimental values (this work): (●), DPA; (■), DBA. Solid lines, calculations with equation (8) using the coefficients from Table 6.

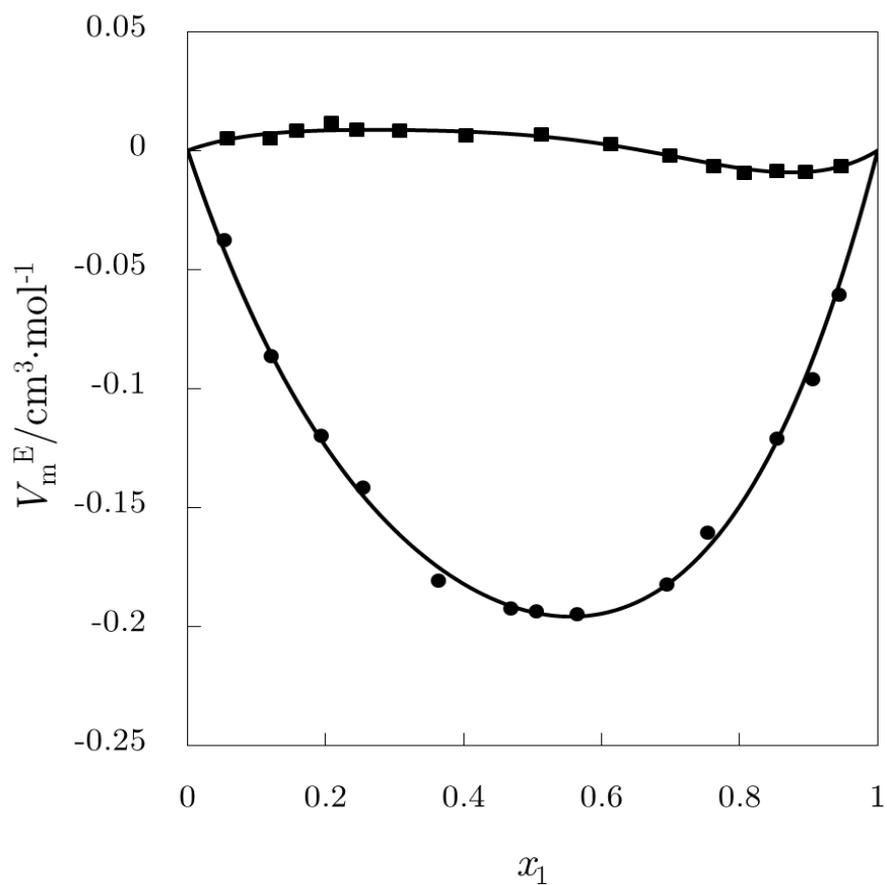

Figure 2

Excess molar volumes, $V_m^E$, for DMA (1) + BA (2), or + HxA (2) systems at 0.1 MPa and 298.15 K. Full symbols, experimental values (this work): (●), BA; (■), HxA. Solid lines, calculations with equation (8) using the coefficients from Table 6.

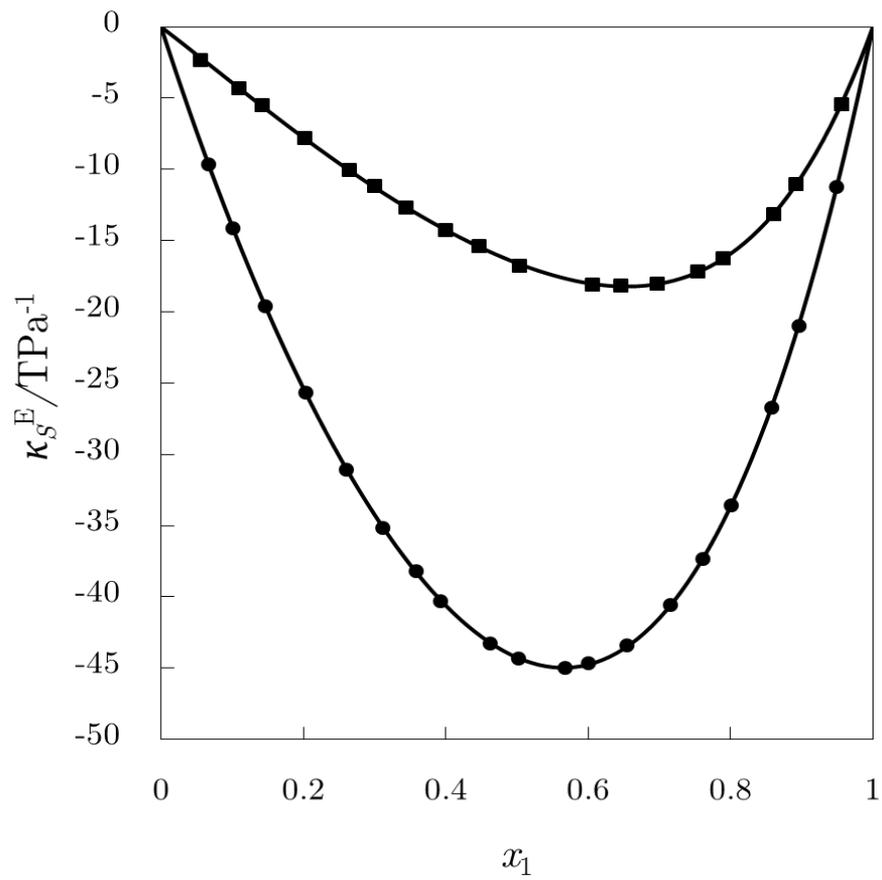

Figure 3

Excess isentropic compressibilities, $\kappa_S^E$, for DMA (1) + DPA (2), or + DBA (2) systems at 0.1 MPa and 298.15 K. Full symbols, experimental values (this work): (●), DPA; (■), DBA. Solid lines, calculations with equation (8) using the coefficients from Table 6.

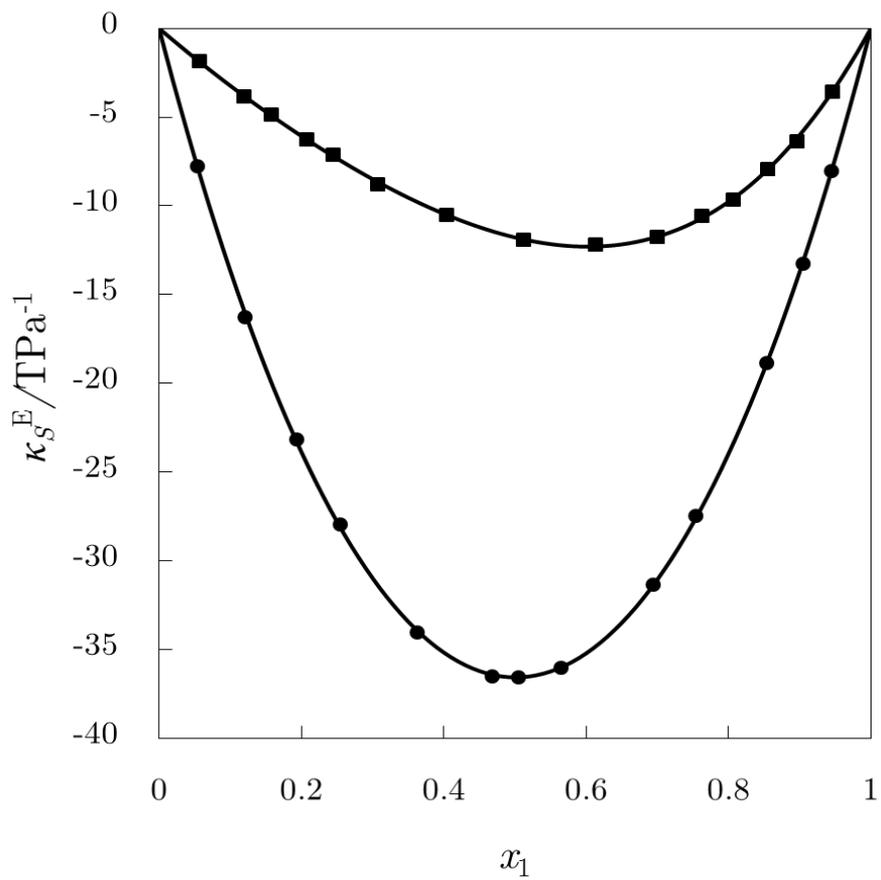

Figure 4

Excess isentropic compressibilities, $\kappa_S^{\mathrm{E}}$, for DMA (1) + BA (2), or + HxA (2) systems at 0.1 MPa and 298.15 K. Full symbols, experimental values (this work): (●), BA; (■), HxA. Solid lines, calculations with equation (8) using the coefficients from Table 6.

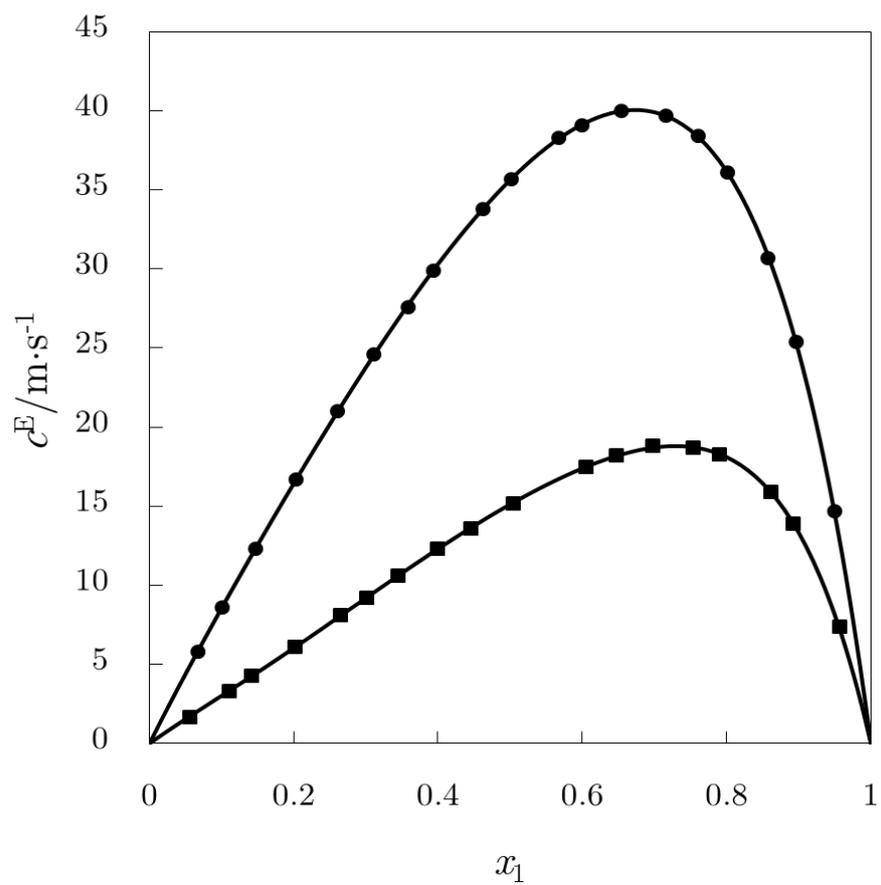

Figure 5

Excess speeds of sound, $c^{\mathrm{E}}$, for DMA (1) + DPA (2), or + DBA (2) systems at 0.1 MPa and 298.15 K. Full symbols, experimental values (this work): (●), DPA; (■), DBA. Solid lines, calculations with equation (8) using the coefficients from Table 6.

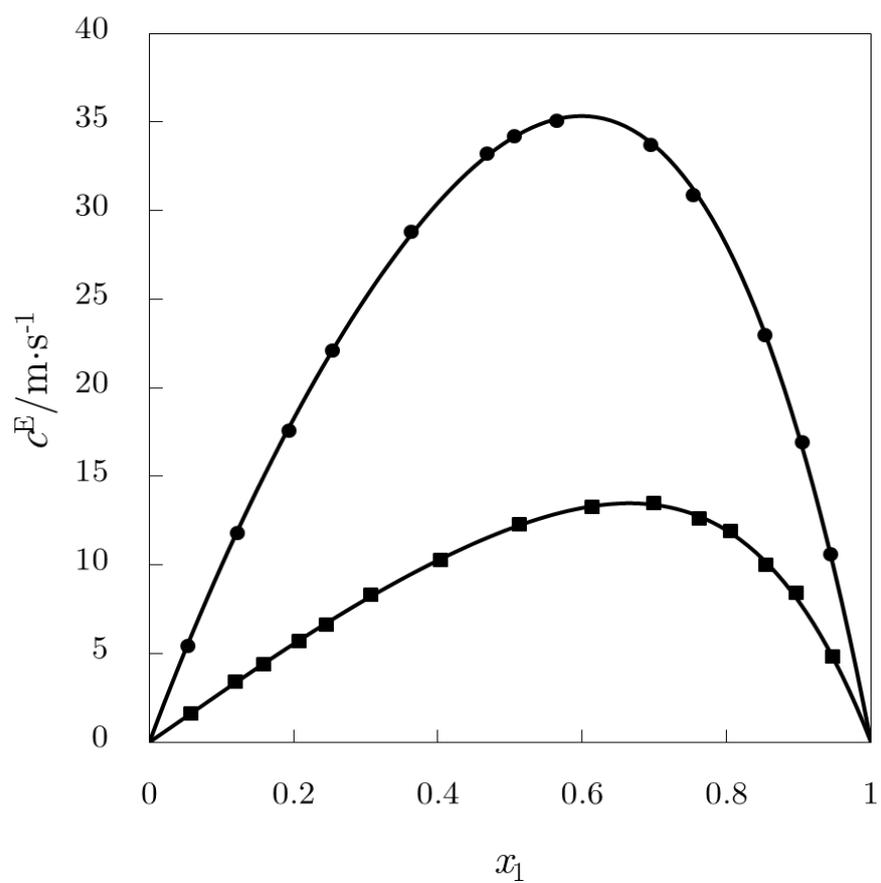

Figure 6

Excess speeds of sound, $c^E$, for DMA (1) + BA (2), or + HxA (2) systems at 0.1 MPa and 298.15 K. Full symbols, experimental values (this work): (●), BA; (■), HxA. Solid lines, calculations with equation (8) using the coefficients from Table 6.

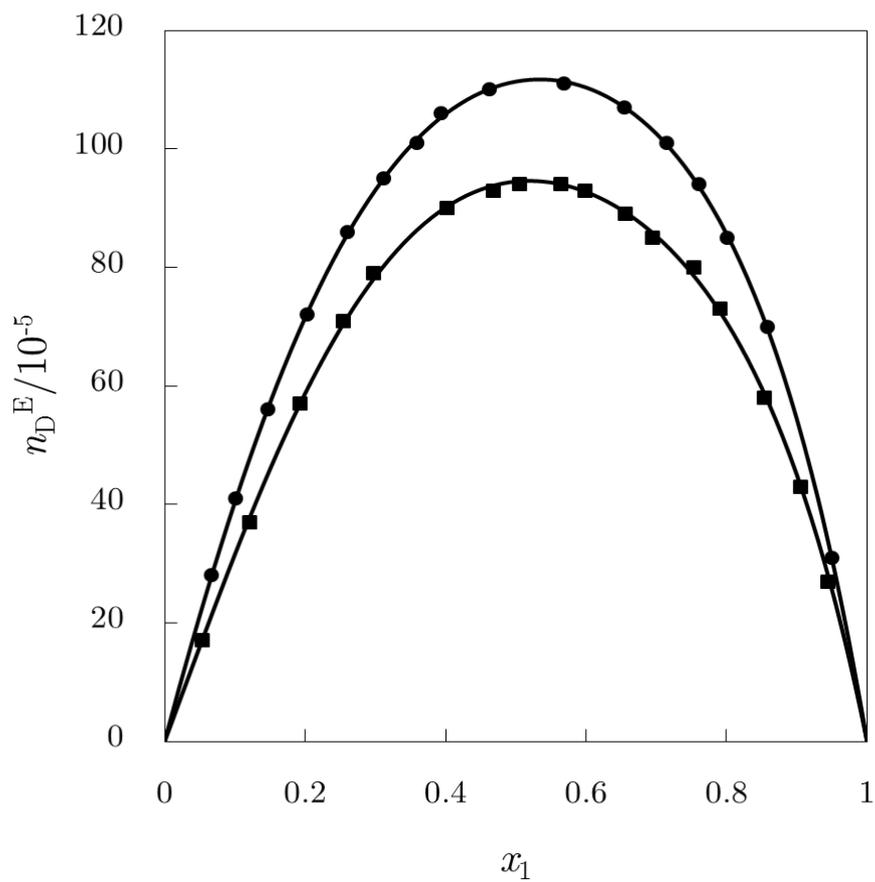

Figure 7

Excess refractive indices, $n_\mathrm{D}^\mathrm{E}$, for DMA (1) + amine (2) systems at 0.1 MPa and 298.15 K. Full symbols, experimental values (this work): (●), DPA; (■), BA. Solid lines, calculations with equation (8) using the coefficients from Table 6.

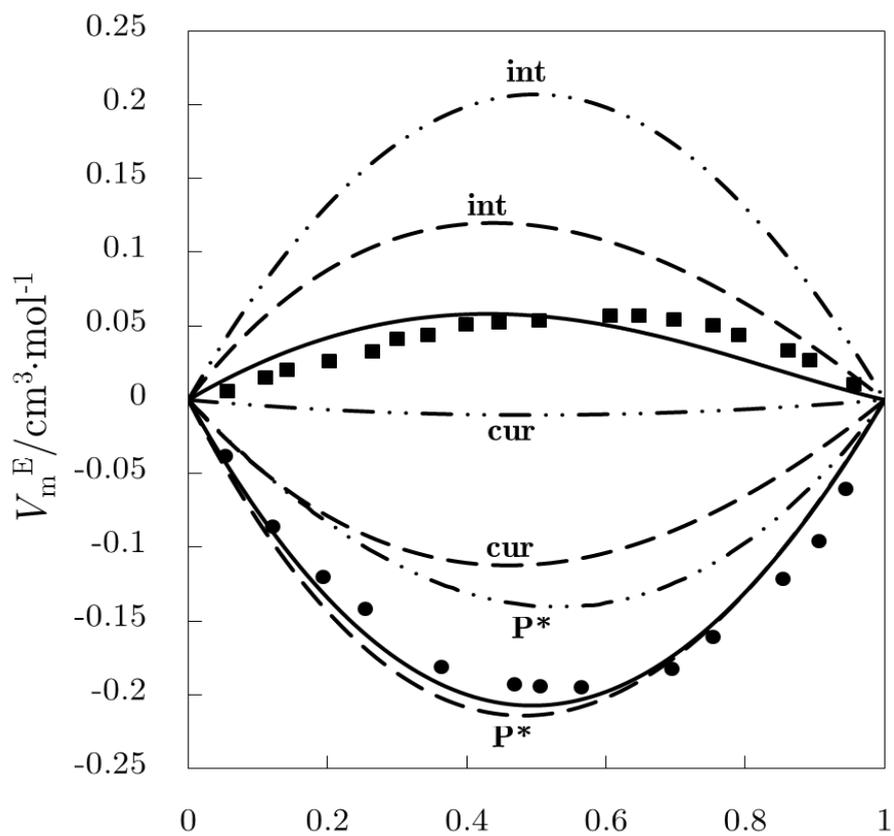

Figure 8

Excess molar volumes, $V_m^E$, for DMA (1) + amine (2) systems at 0.1 MPa and 298.15 K. Full symbols, experimental values (this work): (●), BA; (■), DBA. Solid lines, Flory results. Dashed lines, contributions to $V_m^E$ according to the Prigogine-Flory-Patterson model ("int", interactional; "cur", curvature; "$P^*$", $P^*$ effect): (– – –), BA; (– · · –), DBA.

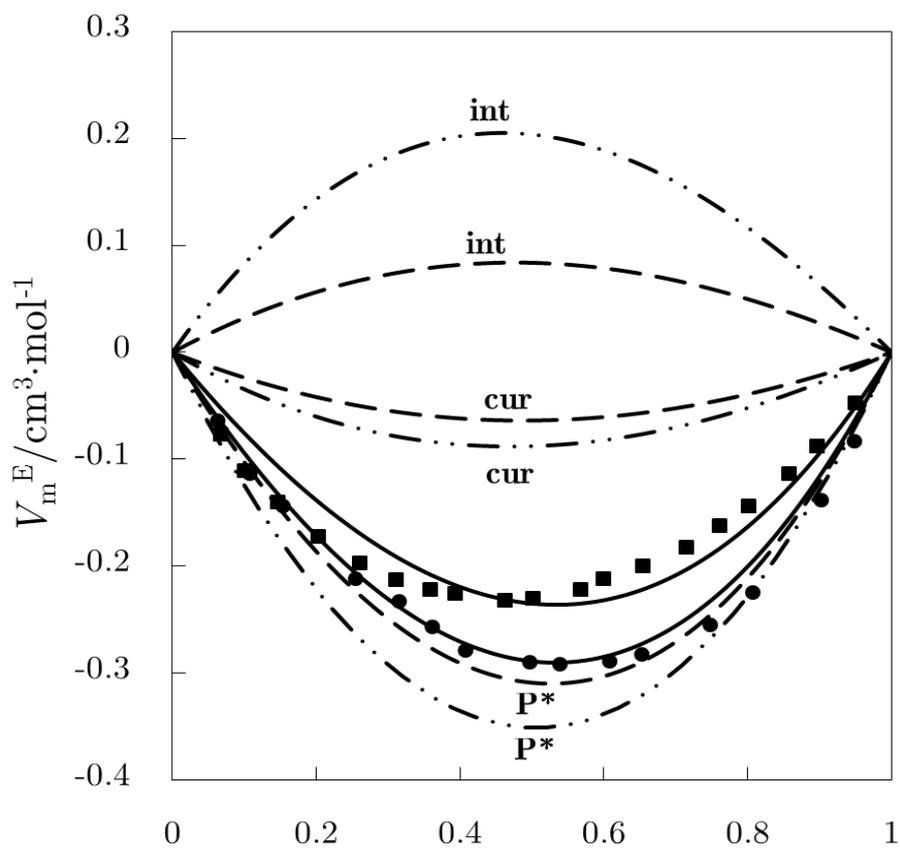

Figure 9

Excess molar volumes, $V_m^E$, for amide (1) + DPA (2) systems at 0.1 MPa and 298.15 K. Full symbols, experimental values : (●), DMF [20]; (■), DMA (this work). Solid lines, Flory results. Dashed lines, contributions to $V_m^E$ according to the Prigogine-Flory-Patterson model ("int", interactional; "cur", curvature; "$P^*$", $P^*$ effect): (– – –), DMF; (– · · –), DMA.